\documentclass[twocolumn,aps,prd,showpacs,10pt,superscriptaddress]{revtex4-1}
\usepackage{amssymb}
\usepackage{amsmath}
\usepackage{multirow}
\usepackage{graphicx}
\usepackage[usenames,dvipsnames,svgnames]{xcolor}
\usepackage{hyperref}
\hypersetup{
pdfnewwindow=true,      
colorlinks=true,        
linkcolor=Blue,         
citecolor=Blue,         
filecolor=Blue,         
urlcolor=Blue           
}

\begin{document}

\title{Heat transport in pristine and polycrystalline single-layer hexagonal boron nitride}

\author{Haikuan Dong}
\affiliation{School of Mathematics and Physics, Bohai University, Jinzhou, China}
\author{Petri Hirvonen}
\affiliation{QTF Centre of Excellence, Department of Applied Physics, Aalto University, FI-00076 Aalto, Finland}
\author{Zheyong Fan}
\email{brucenju@gmail.com}
\affiliation{School of Mathematics and Physics, Bohai University, Jinzhou, China}
\affiliation{QTF Centre of Excellence, Department of Applied Physics, Aalto University, FI-00076 Aalto, Finland}
\author{Tapio Ala-Nissila}
\affiliation{QTF Centre of Excellence, Department of Applied Physics, Aalto University, FI-00076 Aalto, Finland}
\affiliation{Center for Interdisciplinary Mathematical Modeling and Departments of Mathematical Sciences, Loughborough University, Loughborough, Leicestershire LE11 3TU, UK}

\date{\today}

\begin{abstract}
We use a phase field crystal model to generate large-scale bicrystalline and polycrystalline single-layer hexagonal boron nitride (h-BN) samples and employ molecular dynamics (MD) simulations with the Tersoff many-body potential to study their heat transport properties. The Kapitza thermal resistance across individual h-BN grain boundaries is calculated using the inhomogeneous nonequilibrium MD method. The resistance displays strong dependence on the tilt angle, the line tension and the defect density of the grain boundaries. We also calculate the thermal conductivity of pristine h-BN and polycrystalline h-BN with different grain sizes using an efficient homogeneous nonequilibrium MD method. The in-plane and the out-of-plane (flexural) phonons exhibit different grain size scalings of the thermal conductivity in polycrystalline h-BN and the extracted Kapitza conductance is close to that of large-tilt-angle grain boundaries in bicrystals.   
\end{abstract}

\maketitle

\section{Introduction}

Monolayer hexagonal boron nitride (h-BN) \cite{alem2009prb} has been isolated from bulk boron-nitride and could be useful as a two-dimensional (2D) dielectric substrate. Due to the large electronic band gap, phonons are the dominant heat carriers in this material. The relatively high thermal conductivity, combined with the high thermal stability and chemical resistance make h-BN a promising candidate for thermal management applications.

Bulk h-BN is reported to have a basal-plane thermal conductivity of about 390 W/mK at room temperature \cite{sichel1976prb}, which is comparable to that of copper. The room-temperature basal-plane thermal conductivity of suspended few-layer h-BN sheets have also been measured recently: Jo \textit{et al.} \cite{jo2013nl} reported $360$ and $250$ W/mK for $11$-layer and  $5$-layer h-BN samples, Zhou \textit{et al.} \cite{zhou2014nr} reported 
$243$ W/mK ($+37$ W/mK; $-16$ W/mK) for 9-layer h-BN samples and Wang  \textit{et al.} \cite{wang2016sr} reported 
$484$ W/mK ($+141$ W/mK; $-24$ W/mK) for exfoliated bilayer h-BN samples. Laminated h-BN thin films, which can be mass produced, have been measured to have reduced thermal conductivities (20 W/mK reported by Zheng \textit{et al.} \cite{zheng2016twodm} and 14 W/mK by Wang \textit{et al.} \cite{wang2018nanoscale}), but they have nevertheless been demonstrated to be useful for thermal management applications \cite{wang2018nanoscale}.

Experimentally grown samples are usually polycrystalline in nature, containing grain boundaries (GBs) which act as extended defects affecting heat transport. There is so far no experimental work exploring the influence of GBs on the heat transport properties of layered h-BN. Theoretically, it is also a major challenge to obtain realistic large-scale samples for transport studies. Conventional atomistic techniques do not display the multiscale characteristics required for modeling the formation and structure of such defected systems. As an example, related molecular dynamics studies are often initialized with iterative growth and annealing of the grains and the GBs \cite{mortazavi2015sr}, typically resulting in rather irregular defect formations along the GBs. Phase field crystal (PFC) is a family of classical density functional approaches for crystalline matter that bridge the atomistic and mesoscopic length scales and, moreover, give access to long, diffusive time scales. The PFC approach is, therefore, better suited for generating realistic low-stress samples as has been recently demonstrated for graphene \cite{hirvonen2016prb}.

We use here a PFC model to generate the samples and employ large-scale molecular dynamics (MD) simulations with the many-body Tersoff potential \cite{tersoff1989prb,sevik2011prb,lindsay2011prb} to study their thermal transport properties. We first calculate the basal-plane lattice thermal conductivity of pristine h-BN using a highly efficient homogeneous nonequilibrium MD (HNEMD) method \cite{evans1982pla,evans1990book,mandadapu2009jcp,mandadapu2009pre,fan2018submitted,xu2018submitted} and crosscheck the results by the standard equilibrium MD (EMD) method based on the Green-Kubo relation \cite{evans1990book,tuckerman2010book}. Then, we calculate the Kapitza resistance/conductance (also called the thermal boundary resistance/conductance) of individual h-BN GBs and explore possible correlations between the Kapitza resistance and the tilt angle, line tension, and defect density of the GBs. Last, we calculate the thermal conductivity of polycrystalline h-BN samples using the HNEMD method and study the scaling of the thermal conductivity with respect to the average grain size.

 \section{Methods}

\subsection{PFC model for h-BN}

We use a PFC model to generate the bi- and polycrystalline samples for the present MD simulations. PFC models are a family of coarse-grained classical density functional methods for microstructural and elastoplastic modeling of crystalline matter. The main advantage of PFC is the access to diffusive time scales of microstructure evolution that elude, for example, conventional MD simulations. Furthermore, up to mesoscopic systems can be modeled with atomic-level resolution. PFC describes systems in terms of smooth, classical density fields $\psi$. The structures are solved for by minimizing a governing free energy functional $F \left( \psi  \right)$. PFC models can be matched to real materials by choosing $F$ and its parameters such that the same lattice symmetry is reproduced. Additionally, elastic properties, defect energies and diffusion rates, among other properties, can be fitted to those of real systems. \cite{elder2002prl, elder2004pre}

We use here a PFC model \cite{taha2017prl} that incorporates two PFC density fields $\psi_{\rm B}$ and $\psi_{\rm N}$ -- one for each atomic species -- that are coupled together to produce the binary honeycomb lattice structure. We employ the same dynamics and model parameters as reported in Ref. \citenum{taha2017prl}. Evolution of $\psi_{\rm B}$ and $\psi_{\rm N}$ is given by conserved dynamics [$\overline{\psi}_{\rm B} \left( t \right) = \overline{\psi}_{\rm N} \left( t \right) = \overline{\psi}$]
\begin{equation}
\label{equation:pfc-dynamics}
\frac{\partial \psi_i}{\partial t} = \nabla^2 \frac{\delta F}{\delta \psi_i},
\end{equation}
where $i =$ B, N respectively for boron and nitrogen, $t$ is time and $F$ is the free energy
\begin{equation}
\begin{split}
F = \int d\boldsymbol{r} \left( \frac{1}{2} \epsilon \psi_{\rm B}^2 + \frac{1}{2} \psi_{\rm B} \left( 1 + \nabla^2 \right)^2 \psi_{\rm B} + \frac{1}{3} g \psi_{\rm B}^3 + \frac{1}{4} \psi_{\rm B}^4 \right.   \\
+ \frac{1}{2} \epsilon \psi_{\rm N}^2 + \frac{1}{2} \psi_{\rm N} \left( 1 + \nabla^2 \right)^2 \psi_{\rm N} + \frac{1}{3} g \psi_{\rm N}^3 + \frac{1}{4} \psi_{\rm N}^4   \\
\left. + \alpha \psi_{\rm B} \psi_{\rm N} + \frac{1}{2} \beta \left( \psi_{\rm B} \left( 1 + \nabla^2 \right)^2 \psi_{\rm N} + \psi_{\rm N} \left( 1 + \nabla^2 \right)^2 \psi_{\rm B} \right)   \right. \\
\left. + \frac{1}{2} w \left( \psi_{\rm B}^2 \psi_{\rm N} + \psi_{\rm B} \psi_{\rm N}^2 \right) \right), 
\end{split} 
\end{equation}
where $\epsilon = -0.3$, $g = -0.5$, $\alpha = 0.5$, $\beta = 0.02$ and $w = 0.3$. We also set the average density $\overline{\psi} = 0.28$. The first two lines incorporate double well potentials and gradient terms giving rise to spatially oscillatory solutions and to elastic behavior. The last two couple $\psi_{\rm B}$ and $\psi_{\rm N}$ together to yield the correct binary honeycomb structure. We note that this model does not take into account the possible effects of localized charges arising from local excess of either species. Furthermore, with the parameters used, the model is symmetric with respect to the two species, \textit{i.e.}, $F\left(\psi_{\rm B}, \psi_{\rm N}\right) = F\left(\psi_{\rm N}, \psi_{\rm B}\right)$. Further details of the model can be found in Ref. \citenum{taha2017prl}.

\subsection{MD methods for thermal transport}

To study the heat transport properties of the large-scale bicrystalline and polycrystalline h-BN samples generated using the PFC model, we employ classical MD simulations. This is the only feasible method that is at a fully atomistic level. A highly efficient MD code called GPUMD (graphics processing units molecular dynamics) \cite{fan2013cpc,fan2015prb,fan2017cpc,fan2017gpumd} fully implemented on graphics processing units was used to perform all the MD simulations in this work. The HNEMD, EMD, and NEMD methods described below have been implemented in GPUMD. The Tersoff empirical potential \cite{tersoff1989prb} parameterized by Sevik \textit{et al}. \cite{sevik2011prb} is used to describe the interatomic interactions in h-BN. For all the systems, we assume a uniform thickness of $0.335$ nm for monolayer h-BN, which is equal to the interlayer distance of bulk h-BN.

\subsubsection{Thermal conductivity from the HNEMD method}

We use the HNEMD method to calculate the thermal conductivity in both pristine and polycrystalline h-BN. This method was first proposed by Evans \cite{evans1982pla,evans1990book} more than three decades ago in terms of two-body potentials. It was later generalized to a special class of many-body potentials (called the cluster potentials) by Mandadapu \textit{et al}. \cite{mandadapu2009jcp,mandadapu2009pre} and recently to all many-body potentials by some of the current authors \cite{fan2018submitted}.

This method gets its name due to the following two features. First, it is a nonequilibrium MD method and a nonzero heat flux is flowing circularly in the (periodic) transport direction of the system. Second, there is neither an explicit heat source nor a sink in the system; the system is driven into a homogeneous nonequilibrium steady-state by some external force. The homogeneous heat current is generated by adding an external driving force \cite{fan2018submitted}
\begin{equation}
\vec{F}_{i}^{\rm ext}
= \sum_{j \neq i} \left(\frac{\partial U_j}{\partial \vec{r}_{ji}} \otimes \vec{r}_{ij}\right) \cdot \vec{F}_{\rm e}
\end{equation}
on each atom $i$. Here, $U_j$ is the site potential associated with atom $j$, $\vec{r}_{ji} = \vec{r}_i-\vec{r}_j$ is the difference between the positions $\vec{r}_i$ and $\vec{r}_j$, and $\vec{F}_{\rm e}$ is a vector parameter controlling the magnitude and direction of the external driving force. Note that we are studying stable solid systems here and the kinetic term which only matters for fluids is thus excluded. This driving force will be added to the 
interatomic force of atom $i$ related to the many-body potential \cite{fan2015prb}
\begin{equation}
\vec{F}_i^{\rm int} = \sum_{j\neq i} \left(\frac{\partial U_i}{\partial \vec{r}_{ij}} - \frac{\partial U_j}{\partial \vec{r}_{ji}} \right),
\end{equation}
to get the total force $\vec{F}_i^{\rm tot}=\vec{F}_i^{\rm ext}+\vec{F}_i^{\rm int}$. Because $\sum_i \vec{F}_i^{\rm ext} \neq 0$, one needs to zero the total force of the system before integrating the equations of motion. Also, the temperature of the system needs to be controlled when the external driving force is applied. For temperature control, we use the simple velocity rescaling method, and we have confirmed that the results do not depend on the details of the thermostat. Details on the technical issues can be found elsewhere
\cite{evans1982pla,mandadapu2009jcp,fan2018submitted}. 

The thermal conductivity $\kappa(t)$ at a given time $t$ in a given direction is directly proportional to the nonequilibrim ensemble average $\langle J(t) \rangle_{\rm ne}$ of the heat current $J$ generated in that direction:
\begin{equation}
\kappa(t) = \frac{\langle J(t)\rangle_{\rm ne}}{TVF_{\rm e}}.
\end{equation}
Here, $T$ is the temperature, $V$ is the volume, $F_{\rm e}=|\vec{F}_{\rm e}|$ and $J$ is a component of the heat current vector derived in Ref. \citenum{fan2015prb}:
\begin{equation}
\vec{J} = \sum_i \sum_{j\neq i} \left( \vec{r}_{ij} \otimes \frac{\partial U_j}{\partial \vec{r}_{ji}} \right) \cdot \vec{v}_i,
\end{equation}
where $\vec{v}_i$ is the velocity of atom $i$. In practice, the ensemble average is replaced by a time average. 
Following previous works \cite{mandadapu2009jcp,fan2018submitted} we redefine $\kappa(t)$ as a running average
\begin{equation}
\label{equation:hnemd}
\kappa(t)=\frac{1}{t}\int_0^t \frac{\langle J(t')\rangle_{\rm ne}}{TVF_{\rm e}}dt',
\end{equation}
and check how it converges. An important issue is to select an appropriate value of $F_{\rm e}$, which has the dimension of inverse length. This parameter should be small enough such that the system is in the linear response regime. It also needs to be large enough to attain a large signal-to-noise ratio. A rule-of-thumb has been given in Ref. \citenum{fan2018submitted}, which states that when $F_{\rm e} \lambda < 1/10$, where $\lambda$ is the average phonon mean free path, linear response is completely assured. In practice, one can first roughly estimate $\lambda$ and then choose a few values of $F_{\rm e}$ to test. From our tests, we find that $F_{\rm e}=0.1$ $\mu$m$^{-1}$ is sufficiently small for pristine h-BN and $F_{\rm e}=1$ $\mu$m$^{-1}$ is sufficiently small for all the polycrystalline h-BN systems. 

Because there is no heat source or sink in this method, the finite-size effects are very small and can be eliminated by using a relatively small simulation cell. For pristine h-BN, we have tested that a simulation cell of size $25\times 25$ nm$^2$ (with 24 000 atoms) is large enough. The polycrystalline h-BN samples are also large enough such that finite-size effects are negligible. Periodic boundary conditions are applied to the in-plane directions. For pristine h-BN, we first equilibrate the system at $300$ K and zero pressure in the NPT ensemble for $1$ ns and then generate the homogeneous heat current for $10$ ns. The time step for integration is chosen to be $1$ fs and four independent runs were performed. For polycrystalline h-BN, we first equilibrate the system at $1$ K and zero pressure in the NPT ensemble for $0.25$ ns, and then heat up the system to $300$ K during $0.25$ ns. Next, we equilibrate the system at $300$ K and zero pressure for $0.5$ ns in the NPT ensemble and then generate the homogeneous heat current for $0.5$ ns. The time step for integration is chosen to be $0.25$ fs and two independent runs were performed for each sample.

\subsubsection{Thermal conductivity from the EMD method}

We also use the EMD method to crosscheck the HNEMD results for pristine h-BN. The EMD method is based on the Green-Kubo relation \cite{evans1990book,tuckerman2010book}, which expresses the running thermal conductivity $\kappa(t)$ as an integral of the heat current autocorrelation function $\langle J(0)J(t) \rangle_{\rm e}$:
\begin{equation}
\label{equation:gk}
\kappa(t) = \frac{1}{k_{\rm B}T^2V}\int_0^{t} \langle J(0) J(t') \rangle_{\rm e} dt',
\end{equation}
where $t$ is the correlation time and $k_{\rm B}$ is the Boltzmann constant. The equilibrium ensemble average $\langle \rangle_{\rm e}$ is evaluated as an average over different time origins. 

Also the EMD method has small finite-size effects and we thus use the same simulation cell and boundary conditions for pristine h-BN as used in the HNEMD method. We first equilibrate the system at 300 K and zero pressure in the NPT ensemble for $1$ ns and then switch to the NVE ensemble and sample the heat current for $40$ ns. A time step of $1$ fs is used and $36$ independent runs were performed.

\subsubsection{Thermal boundary resistance from the NEMD method}

\begin{figure}[htb]
\centering
\includegraphics[width=\columnwidth]{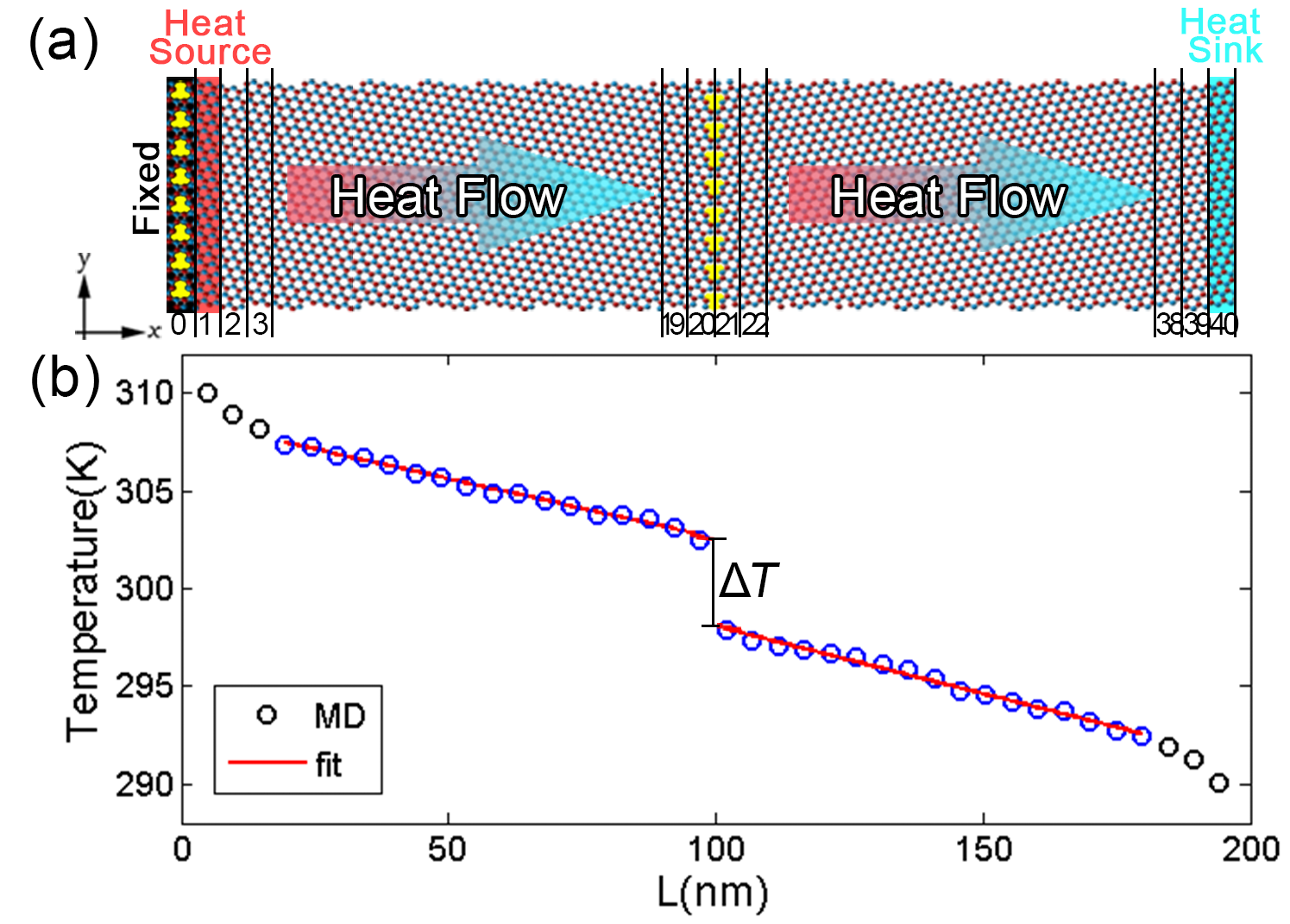}
\caption{(a) A schematic illustration of the NEMD simulation method in bicrystalline h-BN. (b) The temperature profile across bicrystalline h-BN in the steady state of the NEMD simulation.}
\label{fig-exp_nemd}
\end{figure}

We use the NEMD method to calculate the Kapitza thermal resistance across individual grain boundaries. The NEMD method is an inhomogeneous method where a heat source and a sink are generated. There are many techniques to generate these and we select the method of local thermostatting. As shown schematically in Fig. \ref{fig-exp_nemd}(a), we apply periodic boundary conditions in both planar directions, but freeze the atoms in a group (group $0$) containing one grain boundary. This is equivalent to fixing both the left and the right ends of the system. The remaining system is evenly divided into $40$ groups and groups $1$ and $40$ are taken as the heat source and sink, respectively. Their local temperatures are maintained at $310$ K and $290$ K, respectively, using a Nos\'{e}-Hoover chain thermostat \cite{tuckerman2010book} considering the conservation of momentum in the heat source and sink groups. One can also use a periodic boundary setup where the heat source and sink are separated by one half of the simulation cell length, but it has been found \cite{azizi2017cb} that this is less efficient than the fixed boundary setup in obtaining the length-convergent Kapitza thermal resistance, because the periodic boundary setup effectively reduces \cite{xu2018submitted} the system length by a factor of two. We have tested that using a simulation cell length of $200$ nm is enough to obtain length-converged Kapitza thermal resistance for the h-BN grain boundaries. In comparison, it has been found \cite{azizi2017cb} that a simulation cell length of $400$ nm is required to obtain length-converged Kapitza thermal resistance for graphene grain boundaries, which can be understood by noticing that the phonon mean free paths are shorter in h-BN than in graphene.

When steady state is established, a linear temperature profile can be observed on each side of the central grain boundary, as shown by the red lines in Fig. \ref{fig-exp_nemd}(b). Right at the grain boundary, there is a clear discontinuity of the temperature $\Delta T$, which is associated with the Kapitza thermal resistance. We determine the value of $\Delta T$ as the difference between the two fitted linear functions evaluated at the grain boundary. In the steady state, the energy exchange rate $dE/dt$ between the heat source/sink and the thermostat also becomes time independent. From this we get the heat flux $Q$ flowing across the grain boundary:
\begin{equation}
Q = \frac{dE/dt}{S},
\end{equation}
where $S$ is the cross-sectional area of the system perpendicular to the transport direction. From the temperature jump $\Delta T$ and the heat flux $Q$, the Kapitza thermal resistance $R$ is calculated by the definition
\begin{equation}
R = \frac{\Delta T}{Q}.
\end{equation}
The inverse of the Kapitza thermal resistance is called the thermal boundary conductance $G$ (here, $G$ is actually the thermal boundary conductance per unit area):
\begin{equation}
G = \frac{1}{R} = \frac{Q}{\Delta T}.
\end{equation}

In the NEMD simulations, we use a time step of $0.25$ fs. We first equilibrate the system at 1 K and zero pressure in the NPT ensemble for $0.25$ ns, then increase the temperature form $1$ K to $300$ K during $0.25$ ns, and then equilibrate the system at 300 K and zero pressure in the NPT ensemble for $0.5$ ns. After these steps, we apply Nos\'{e}-Hoover chain  thermostats locally to the heat source and sink, setting the local temperature to $310$ K and $290$ K, respectively. We have checked that all the systems can reach a steady state during $0.5$ ns. In view of this, the local thermostats are applied for $1.5$ ns and we measure the heat flux and the temperature profile during the last $1$ ns. We do five independent runs for each sample in order to obtain an average and an error estimate for each physical quantity we are interested in. 

\section{Results and Discussion}

\subsection{Sample preparation and inspection}

\begin{figure*}
\includegraphics[width=\textwidth]{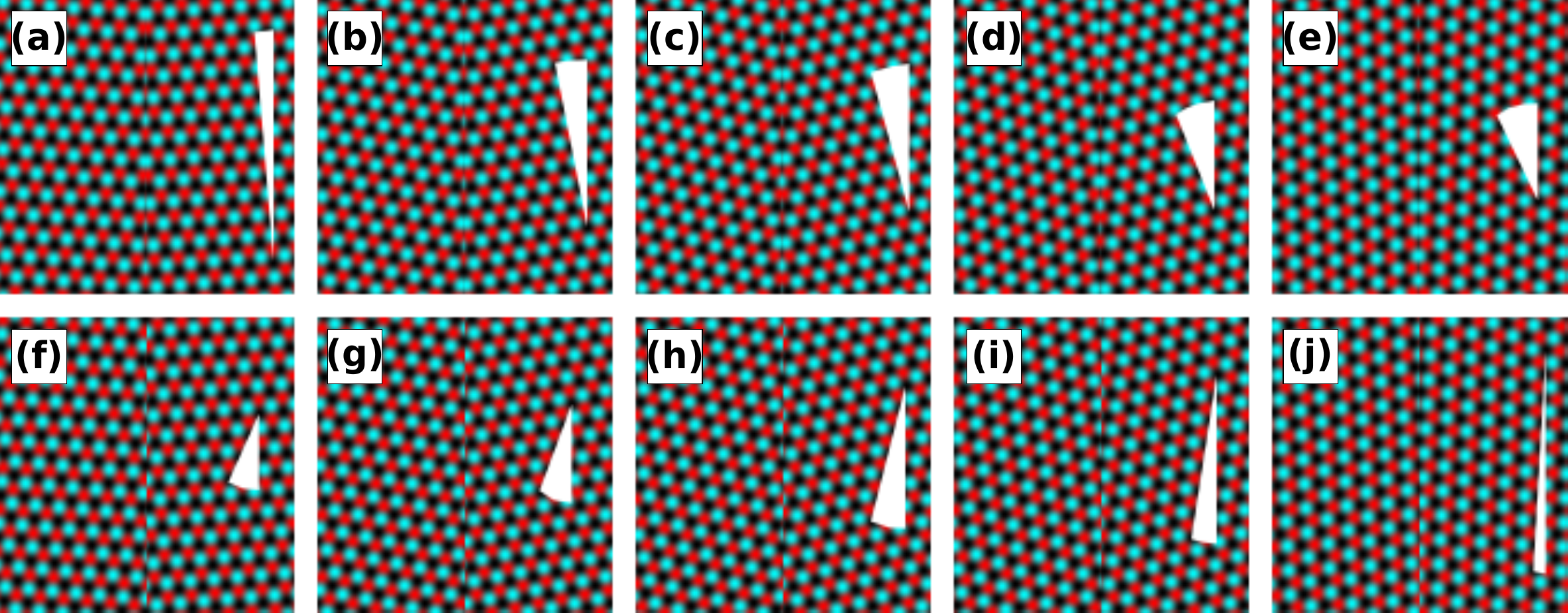}
\caption{Collage of some of the grain boundary tilt angle cases studied. Only small portions of the systems are illustrated. The boundaries are shown unrelaxed with sharp interfaces to better illustrate the two symmetry classes: (a) - (e) symmetric and (f) - (j) antisymmetric boundaries. The latter are otherwise identical to the former, but the two atom species in the left-hand side grain of the latter have been interchanged. The wedges indicate one half of the tilt angle, defined as rotation from the single-crystalline state. The tilt angles in (a) - (j) are $9.4^\circ, 21.8^\circ, 32.2^\circ, 42.1^\circ, 53.6^\circ, 50.6^\circ, 38.2^\circ, 27.8^\circ, 17.9^\circ$ and $6.4^\circ$, respectively.}
\label{fig-collage}
\end{figure*}

Similarly to Ref. \citenum{azizi2017cb}, we initialize the bicrystal PFC calculations with symmetrically tilted grains fitted to a periodic, two-dimensional computational unit cell whose dimensions are allowed to vary to minimize strain. The grain boundaries are initialized as sharp interfaces. Figure \ref{fig-collage} presents a collage of the initial grain boundary configurations for some of the tilt angle cases studied.

We consider the same tilt angles as in the previous work \cite{azizi2017cb} for symmetric boundaries (SBs) in graphene. In addition, we also investigate the antisymmetric boundaries (ABs) where the two atomic species have been interchanged in just one of the two grains; compare Figs. \ref{fig-collage} (f)-(j) to (a)-(e). We refer to both boundary types collectively as grain boundaries (GBs). We define the tilt angle as $2\theta$ where $\pm\theta$ is the rotation angle of the grains from the single-crystalline state. Small (large) $2\theta$ corresponds to armchair and zigzag (zigzag and armchair) boundaries for SBs and ABs, respectively; see Fig. \ref{fig-collage}.

After equilibration using Eq. (\ref{equation:pfc-dynamics}), the local maxima in $\psi_{\rm B}$ and $\psi_{\rm N}$, whose magnitude $\geq 75\%$ of that of the global maximum, are detected and their locations are recorded as atom coordinates for further atomistic calculations. We find that the network of atoms, or the topology of the PFC samples remains unchanged during our MD simulations. 

In Ref. \citenum{taha2017prl} it is noted that the relative lattice translation of the neighboring grains in the initial state was varied to find the lowest-energy GB structures. Our goal, however, is to obtain realistic configurations for further atomistic calculations without extensive sampling. For polycrystalline large-scale samples, varying the translations of each grain is infeasible. We consider only a single sample per each tilt angle case and work with the boundaries the model gives.

\begin{table*}
\centering
\caption{Overview of the h-BN grain boundaries from the present (corresponding results without references) and previous \cite{liu2012an, ding2014pccp, taha2017prl} works. The first column gives the tilt angle, whereas the following ones indicate whether armchair (AC) or zigzag (ZZ) boundaries are in question and what defects comprise the corresponding SBs and ABs. Expected ground state configurations are given in boldface. The subscript "B" indicates a boron-rich boundary variant. Alternative structures, whose energies are comparable to the ground state configurations, are given in parentheses. Notation $\left(P|\right)_n$ indicates a defect chain of $n$ $P$-gons. At $2\theta = 0^\circ$ and $2\theta = 60^\circ$ a single-crystalline state and inversion boundaries (IBs) are obtained, respectively. The former case is trivial, whereas for the latter several alternative configurations have been proposed. The related acronym "PH-HEB" stands for "pristine hexagonal with homoelemental bonds".}
\label{tab-results}
\begin{tabular}{ccccc}
\hline
\hline
Tilt angle & AC/ZZ & SBs & AC/ZZ & ABs \\
\hline

$2\theta = 0^\circ$ & AC & \textbf{single-crystalline} & ZZ & \textbf{single-crystalline} \\

\hline

$0^\circ < 2\theta < 30^\circ$ & AC & \textbf{5\textbar7$_\textrm{B}$} (4\textbar8) \cite{liu2012an} & ZZ & \textbf{4\textbar8} (5\textbar7) \cite{liu2012an} \\

 &  & \textbf{5\textbar7} \cite{taha2017prl} &  &  \textbf{5\textbar7}, \textbf{4}, \textbf{12} \cite{taha2017prl} \\

 &  &  4, 5\textbar7, 8 &  &  4, 5\textbar7, 8, 12, 4\textbar12 \\

\hline

$30^\circ < 2\theta < 60^\circ$ & ZZ & \textbf{5\textbar7$_\textrm{B}$} \cite{liu2012an} & AC & \textbf{4\textbar8} \cite{liu2012an} \\

 &  & \textbf{5\textbar7} \cite{taha2017prl} &  & \textbf{5\textbar7}, \textbf{8}, \textbf{4\textbar12} \cite{taha2017prl} \\

 &  & $\left(4|\right)_n$ 5\textbar7 $\left(|8\right)_m$ &  & 4\textbar8, 4\textbar12, 12 \\

\hline

$2\theta = 60^\circ$ & ZZ & \textbf{PH-HEB} \cite{liu2012an} & AC & \textbf{4\textbar8} \cite{liu2012an} \\

 &  & PH-HEB \cite{ding2014pccp} &  & PH-HEB \cite{ding2014pccp} \\

 &  & \textbf{8\textbar8} (4\textbar4) \cite{taha2017prl} &  & \textbf{4\textbar8} \cite{taha2017prl} \\

\hline
\hline
\end{tabular}
\end{table*}

Both atomistic calculations \cite{liu2012an} and experiments \cite{gibb2013jacs, li2015nl} suggest that h-BN SBs and ABs are composed mainly of 5|7 and 4|8, or pentagon-heptagon and tetragon-octagon dislocations, respectively. The h-BN PFC model has been shown to capture many of the correct GB structures \cite{taha2017prl}, but also some alternative ones, including diamond and octagon dislocations, seldom also dodecagon and 4\textbar12 dislocations (interpreted as 4\textbar10 and 4\textbar6\textbar8 in Ref. \citenum{taha2017prl}).

\begin{figure}[htb]
\centering
\includegraphics[width=\columnwidth]{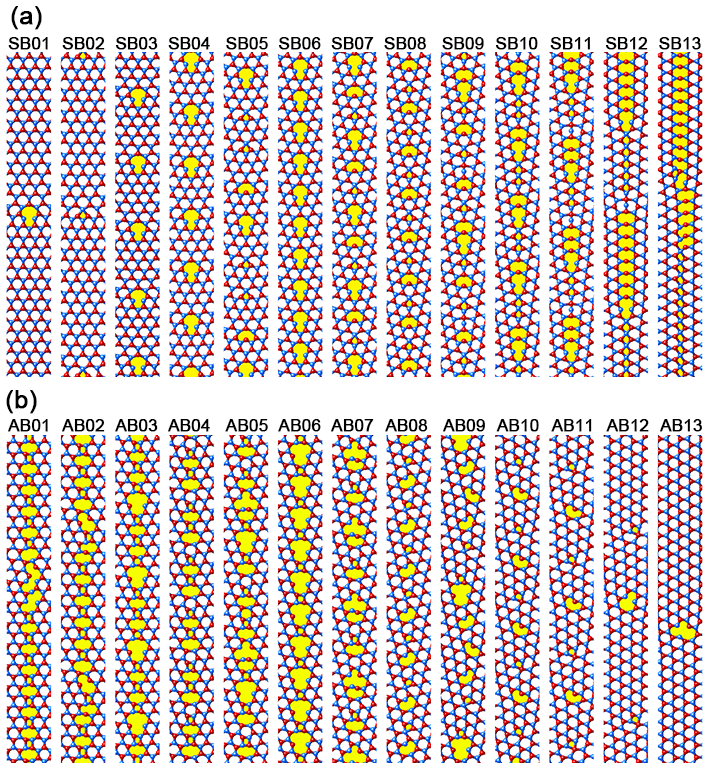}
\caption{Atomistic structures of (a) SBs and (b) ABs before MD relaxation. The defect distributions of the systems remain unchanged during the MD simulations. The tilt angles of the systems are given in Table \ref{tab_02}.}
\label{fig-GBs}
\end{figure}

Figure \ref{fig-GBs} illustrates GB structures from present PFC calculations and Table \ref{tab-results} summarizes and compares them with those from previous theoretical works \cite{liu2012an, ding2014pccp, taha2017prl}. Experimental works \cite{gibb2013jacs, li2015nl} are excluded from Table \ref{tab-results} because they offer a very limited and non-representative sample. Our GBs exhibit an abundance of the 5|7 and 4|8 dislocations as expected \cite{liu2012an, gibb2013jacs, li2015nl}, but display also all of the other dislocation types mentioned above, in accordance with Ref. \citenum{taha2017prl}. Short chains of 5\textbar7 dislocations have been observed experimentally \cite{gibb2013jacs} in GBs of misorientation \textasciitilde $21^\circ$ and \textasciitilde $22^\circ$. Our SB06 structure with a similar misorientation is consistent and also displays 5\textbar7 dislocations. Chains of 4\textbar8 dislocations have been observed in GBs of misorientation \textasciitilde $24^\circ$. Our AB06 and AB07 structures have roughly similar misorientations and do display both squares and octagons, but also dodecagons in the former which have not been observed experimentally or been predicted by atomistic calculations. While the dodecagons form a very regular structure in this case, they are tiny voids that impede traversing phonons. This is evident in Fig \ref{fig_kr} where the corresponding resistance is slightly higher than those of the other GBs.

Chains of 5\textbar7 dislocations have also been observed \cite{li2015nl} in GBs of misorientation \textasciitilde $32^\circ$. Our SB07, SB08 AB07 and AB08 structures all have the same misorientation. Structure SB07 is composed of a mixture of 5|7, diamond and octagon dislocations, structures SB08 and AB08 display diamonds and octagons, and structure AB07 has squares, octagons and dodecagons as discussed earlier. We would like to point out that separate diamonds and octagons are rather similar to 5\textbar7 dislocations, as a diamond (an octagon), together with a neighboring hexagon, can transform into a 5\textbar7 with the introduction (removal) of an atom; see, for example, structure SB07 in Fig. \ref{fig-GBs} that demonstrates all three dislocation types. Diamonds and octagons have also been observed in a material that is structurally quite similar, namely in molybdenum disulfide \cite{zhou2013nl}.

The PFC model rarely produces separate 4\textbar8 dislocations (\textit{e.g.}, in structure AB07 in Fig. \ref{fig-GBs}) that, based on atomistic calculations \cite{liu2012an}, would be expected in zigzag ABs where $2\theta < 30^\circ$. However, 4\textbar8 dislocations are reportedly favored only slightly over 5\textbar7 dislocations in terms of their formation energy whereby the true preference might not be so clear here.

Lastly, a tilt angle of $2\theta = 60^\circ$ corresponds to an inversion boundary between two grains where the two atomic species are swapped when moving from one grain to the other. A defect line of homoelemental bonds with otherwise pristine honeycomb lattice is expected for SBs \cite{liu2012an}, but the h-BN PFC model appears to prefer chains of diamonds and octagons. As a consequence, zigzag SBs where $2\theta > 30^\circ$ have partially incorrect structures; see the structures SB10 - SB13 in Fig. \ref{fig-GBs}, for example. It is then possible that for zigzag SBs the heat transport results may be somewhat deviant.

\begin{figure}[htb]
\centering
\includegraphics[width=\columnwidth]{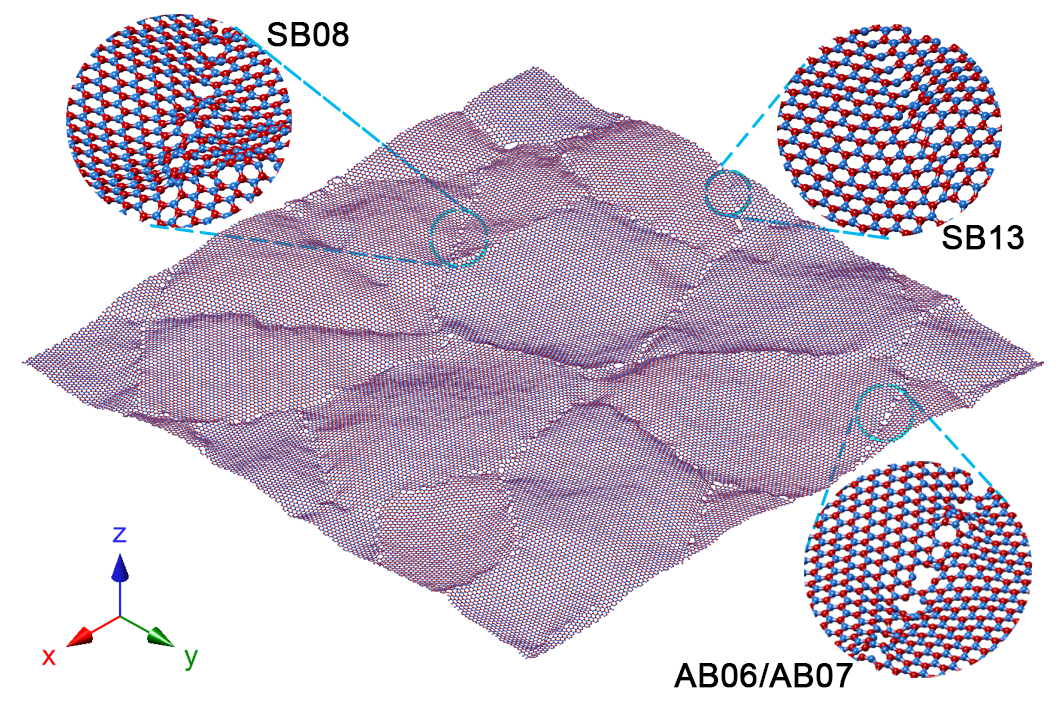}
\caption{Atomistic structure of a polycrystalline h-BN sample after the MD simulation. This is a sample with the smallest grain size ($10$ nm) and sample size ($40\times40$ nm$^2$). The GB structures in the random polycrystalline samples are largely similar to the (anti)symmetrically tilted bicrystal GBs. A few examples are showcased and the corresponding similar
bicrystal GBs are indicated; \textit{cf.} Fig. \ref{fig-GBs}.}
\label{fig-polycry}
\end{figure}

The polycrystalline samples are constructed using random Voronoi tessellations of h-BN crystals relaxed with the PFC model. The resulting networks of grains and grain boundaries are similar to those in polycrystalline samples by CVD \cite{gibb2013jacs, li2015nl}. These GBs display rather regular arrays of dislocations, like their bicrystal counterparts, and lack exotic clusters of nonhexagons often found in samples prepared by iterative growth and annealing; \textit{cf.} Ref. \citenum{mortazavi2015sr}, for example. Figure \ref{fig-polycry} shows an example of our samples, where typical GBs are highlighted.

For polycrystalline samples, we study the influence of the average grain size $d$ on heat transport. We define the average grain size as
\begin{equation}
d = \sqrt{\frac{A}{N}},
\end{equation}
where $A$ and $N$ are the $xy$-projected area of the sample and the number of grains comprising it, respectively. We construct polycrystalline h-BN samples with six grain sizes: $d=10$ nm, $12.5$ nm, $17.5$ nm, $25$ nm, $37.5$ nm, and $50$ nm. All the samples are almost square shaped with the linear size $L_x=L_y=4d$. For each grain size, we construct eight samples.

\subsection{Thermal conductivity of pristine h-BN}

\begin{figure}[htb]
\centering
\includegraphics[width=1\linewidth]{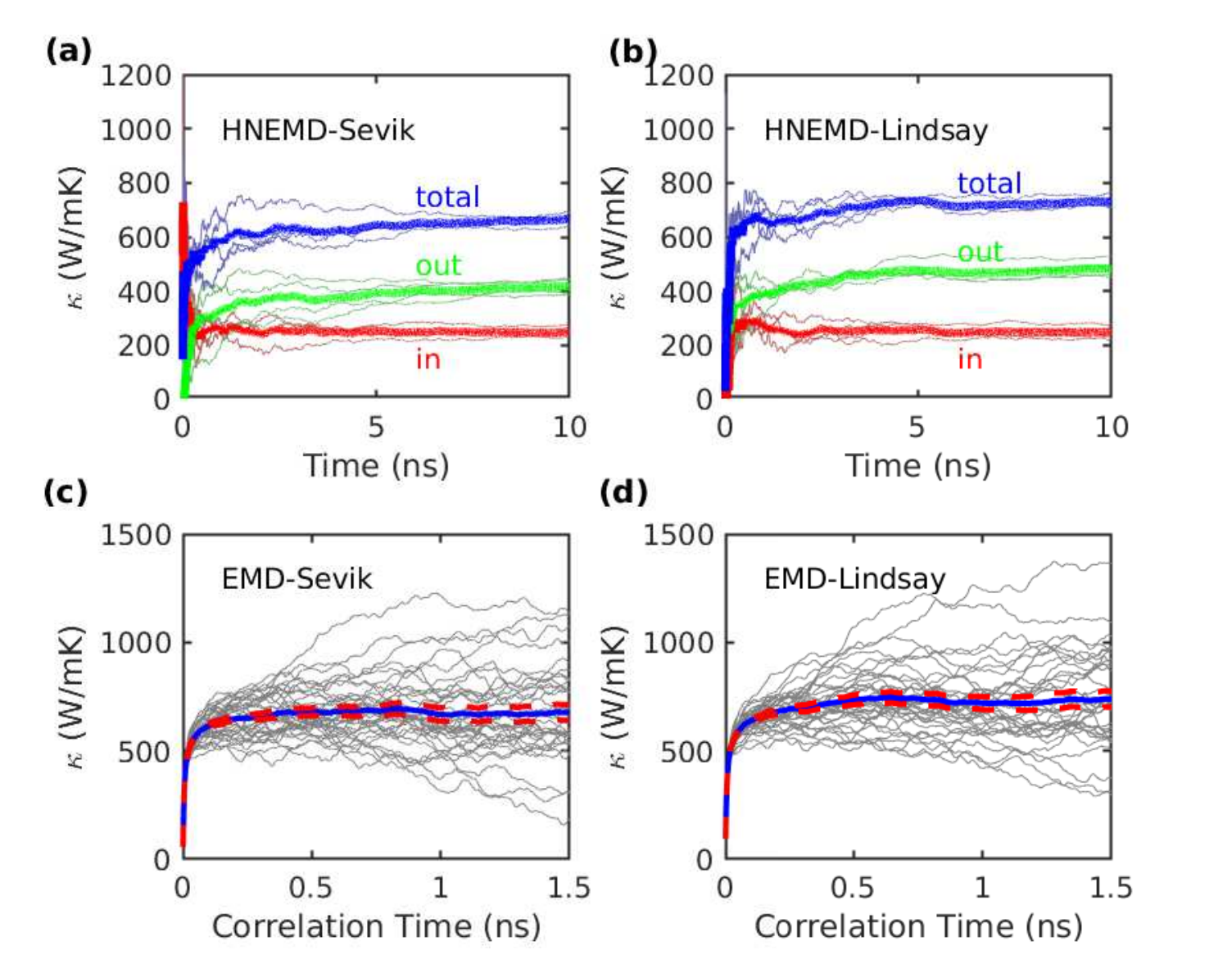}
\caption{(a,b) Running average of the thermal conductivity $\kappa$ as a function of time $t$ calculated using the HNEMD method for h-BN at 300 K. The total thermal conductivity (labeled as ``total'') is decomposed into an in-plane component (labeled as ``in'') and an out-of-plane component (labeled as ``out''). The thin lines are from four independent runs and the thick lines are the mean values. (c,d) Running thermal conductivity $\kappa$ as a function of the correlation time $t$ calculated using the EMD method for pristine h-BN at 300 K. The thick solid line is an average over the $36$ independent runs (each with a production time of $40$ ns) represented by the thin lines and the thick dashed lines indicate the running standard error (standard deviation divided by the square root of the number of independent runs). The Tersoff parameterizations are indicted in the plots.}
\label{fig_pristine}
\end{figure}

Figures \ref{fig_pristine}(a) and (b) show the running average of the thermal conductivity defined in Eq. (\ref{equation:hnemd}), using the Tersoff parameterizations from Sevik \textit{et al.} \cite{sevik2011prb} and Lindsay and Broido \cite{lindsay2011prb}, respectively. Following Ref. \citenum{fan2017prb}, we decompose the heat current into an in-plane component and an out-of-plane one as
\begin{equation}
\vec{J}^{\rm in} = \sum_i  \sum_{j\neq i} \vec{r}_{ij} 
\left(\frac{\partial U_j}{\partial x_{ji}} v^{x}_{i}+
\frac{\partial U_j}{\partial y_{ji}} v^{y}_{i}\right);
\end{equation}
\begin{equation}
\vec{J}^{\rm out} = \sum_i \sum_{j\neq i} \vec{r}_{ij} \frac{\partial U_j}{\partial z_{ji}} v^{z}_{i},
\end{equation}
corresponding to the in-plane and out-of-plane phonons, respectively. In this way, the thermal conductivity is also decomposed into two components, $\kappa_0^{\rm in}$ and $\kappa_0^{\rm out}$. Here, we use a subscript ``0'' to indicate the results for pristine h-BN. With increasing time, the running average of the thermal conductivity becomes more and more stable. Based on the four independent runs up to $10$ ns, we obtain $\kappa_0^{\rm in}=250\pm 10$ W/mK, $\kappa_0^{\rm out}=420\pm 20$ W/mK, and $\kappa_0 = \kappa_0^{\rm in} + \kappa_0^{\rm out} = 670 \pm 30$ W/mK for the parameterization in Ref. \citenum{sevik2011prb}. Using the parameterization in Ref. \citenum{lindsay2011prb}, we get $\kappa_0^{\rm in}=250\pm 10$ W/mK, $\kappa_0^{\rm out}=480\pm 20$ W/mK, and $\kappa_0 = \kappa_0^{\rm in} + \kappa_0^{\rm out} = 730 \pm 30$ W/mK.

To crosscheck the HNEMD results, we also calculate the thermal conductivity of pristine h-BN using the EMD method. Figures \ref{fig_pristine} (c) and (d) show the running thermal conductivity from the EMD simulations using the two Tersoff parameterizations. We see that the averaged running thermal conductivity converges well up to a correlation time of $1.5$ ns. Based on the $36$ independent runs, we get $\kappa_0=670\pm30$ W/mK using the parameterization in Ref. \citenum{sevik2011prb} and $\kappa_0 =730 \pm 30$ W/mK using the parameterization in Ref. \citenum{lindsay2011prb}, which are in excellent agreement with the HNEMD results. We note that each run in the EMD simulations lasts $40$ ns in the production stage, while each run in the HNEMD simulations only lasts $10$ ns. We see that the HNEMD method is more than an order of magnitude faster than the EMD method, as has also been demonstrated for other materials \cite{fan2018submitted,xu2018submitted}.

Using the parameterization in Ref. \citenum{sevik2011prb}, our predicted thermal conductivity value  ($\kappa_0=670\pm30$ W/mK) is significantly higher than that ($\kappa_0=400$ W/mK) calculated by Sevik \textit{et al.} \cite{sevik2011prb} using an Einstein relation \cite{kinaci2012jcp}, which was claimed to be equivalent to the Green-Kubo relation in the EMD method. However, we note that the energy moment formulas in Ref. \citenum{kinaci2012jcp} were not rigorously derived and the calculated thermal conductivity for a silicon crystal at 300 K using the Tersoff potential \cite{tersoff1989prb} ($160.5\pm 10.0$ W/mK) is also significantly smaller than the value ($250\pm 10$ W/mK) obtained by some of the current authors \cite{dong2018prb,fan2018submitted} using various MD based methods. Using the other parameterization \cite{lindsay2011prb} of the Tersoff potential \cite{tersoff1989prb}, Mortazavi \textit{et al.} \cite{mortazavi2015sr} obtained a value of $300\pm30$ W/mK for pristine h-BN, which is significantly smaller than our predicted value of $730\pm30$ W/mK. The smaller value reported in Ref. \citenum{mortazavi2015sr} is due to the wrong heat current formula for many-body potentials as implemented in the LAMMPS code \cite{plimpton1995jcp,lammps} used in their work. It has been emphasized \cite{fan2015prb,fan2017prb,dong2018prb,xu2018submitted} that in these cases the heat current implemented in the LAMMPS code consistently underestimates the thermal conductivity of 2D materials. 

The results above are for isotopically pure systems. While natural nitrogen consists of mainly ($>99.6\%$) $^{14}$N, natural boron consists of $20\%$ $^{10}$B and $80\%$ $^{11}$B. The large concentration of $20\%$ $^{10}$B will induce strong isotopic impurity  scatterings. Using the Tersoff parameterization in Ref. \citenum{sevik2011prb} and considering random distributions of $^{10}$B and $^{11}$B, we get a converged thermal conductivity of $\kappa=520 \pm 30$ W/mK. That is, isotopically pure h-BN has about $30\%$ larger thermal conductivity as compared to the naturally occurring  h-BN, which is in agreement with the prediction using perturbation theory in the Boltzmann transport equation approach \cite{lindsay2011prb}. Our predicted $\kappa$ for single-layer naturally occurring h-BN is above but close to the measured value ($484^{+141}_{-24}$ W/mK) for exfoliated bilayer h-BN samples \cite{wang2016sr}, indicating that the empirical potential \cite{sevik2011prb} is reliable. Phonon dispersion of pristine h-BN calculated using this potential \cite{sevik2011prb} agrees very well with experimental data. In the following, for simplicity, we chose to use the Tersoff parameterization in Ref. \citenum{sevik2011prb} to study heat transport in isotopically pure bi- and polycrystalline h-BN systems.

\subsection{Kapitza thermal resistance across individual h-BN grain boundaries}

\begin{table*}[htb]
\centering
\caption{The tilt angle $2\theta$, temperature jump $\Delta T$, heat flux $Q$, Kapitza resistance $R$, Kapitza length $L_{\rm K}$, grain boundary line tension $\gamma$, thermal boundary conductance $G$, and defect density $\rho$ for the 26 bicrystalline h-BN samples (13 SBs and 13 ABs) used in the NEMD simulations.}
\label{tab_02}
\begin{tabular}{crrrrrrrr}
\hline
\hline
sample & $2\theta$ ($^\circ$) & $\Delta T$ (K)  & $Q$ (GW/m$^2$)  & $R$ (m$^2$K/GW)  & $L_{\rm K}$ (nm)  & $\gamma$ (eV/nm)  & $G$ (GW/m$^2$K) & $\rho$ (1/nm)\\
\hline
SB01    & 1.10    & 0.21$\pm$0.05     & 34.2$\pm$0.3    & 0.0061$\pm$0.0015 & 4.1$\pm$1.0       & 0.534$\pm$0.001      & 296.5 $\pm$125.7     & 0.077 \\
SB02    & 4.41    & 0.63$\pm$0.04     & 33.0$\pm$0.2    & 0.0189$\pm$0.0012 & 12.7$\pm$0.8       & 2.253$\pm$0.020      & 53.9 $\pm$3.3       & 0.307\\
SB03    & 9.43    & 1.11$\pm$0.10     & 33.0$\pm$0.6    & 0.0336$\pm$0.0029 & 22.5$\pm$1.9       & 3.863$\pm$0.022      & 31.0 $\pm$2.9       & 0.657 \\
SB04    & 13.17   & 1.45$\pm$0.03     & 32.7$\pm$0.4    & 0.0443$\pm$0.0008 & 29.7$\pm$0.5       & 5.001$\pm$0.011      & 22.6 $\pm$0.4       & 0.920 \\
SB05    & 18.73   & 2.47$\pm$0.08     & 30.9$\pm$0.4    & 0.0799$\pm$0.0026 & 53.5$\pm$1.8       & 6.714$\pm$0.017      & 12.6 $\pm$0.4       & 1.315\\
SB06    & 21.79   & 1.65$\pm$0.11     & 32.6$\pm$0.4    & 0.0508$\pm$0.0038 & 34.0$\pm$2.5       & 6.813$\pm$0.015      & 20.3 $\pm$1.5       & 1.534 \\
SB07    & 27.80   & 2.79$\pm$0.14     & 29.9$\pm$0.3    & 0.0933$\pm$0.0047 & 62.5$\pm$3.2       & 8.272$\pm$0.011      & 10.8 $\pm$0.5       & 1.972 \\
SB08    & 32.20   & 3.32$\pm$0.09     & 28.1$\pm$0.3    & 0.1182$\pm$0.0021 & 79.2$\pm$1.4       & 9.170$\pm$0.005      & 8.5  $\pm$0.1       & 1.972 \\
SB09    & 36.52   & 3.20$\pm$0.12     & 28.2$\pm$0.2    & 0.1179$\pm$0.0057 & 79.0$\pm$3.8       & 9.924$\pm$0.011      & 8.6  $\pm$0.5       & 1.657\\
SB10    & 42.10   & 2.81$\pm$0.17     & 30.3$\pm$0.6    & 0.0926$\pm$0.0046 & 62.0$\pm$3.1       & 9.307$\pm$0.039      & 11.0 $\pm$0.6       & 1.255 \\
SB11    & 46.83   & 3.17$\pm$0.10     & 27.8$\pm$0.3    & 0.1140$\pm$0.0031 & 76.4$\pm$2.1       & 10.261$\pm$0.013     & 8.8 $\pm$0.2        & 0.920 \\
SB12    & 53.60   & 2.63$\pm$0.07     & 29.9$\pm$0.3    & 0.0878$\pm$0.0022 & 58.8$\pm$1.5       & 8.667$\pm$0.006      & 11.4 $\pm$0.3       & 0.445\\
SB13    & 59.04   & 2.58$\pm$0.04     & 29.4$\pm$0.3    & 0.0879$\pm$0.0011 & 58.9$\pm$0.7       & 7.114$\pm$0.001      & 11.4 $\pm$0.1       & 0.067 \\
AB01    & 58.90    & 3.02$\pm$0.16     & 29.3$\pm$0.2    & 0.1031$\pm$0.0055 & 69.1$\pm$3.7       & 7.546$\pm$0.013      & 9.8  $\pm$0.5      & 0.077\\
AB02    & 55.59    & 2.70$\pm$0.11     & 28.9$\pm$0.4    & 0.0937$\pm$0.0033 & 62.8$\pm$2.2       & 7.938$\pm$0.012      & 10.7 $\pm$0.4      & 0.307  \\
AB03    & 50.57    & 3.55$\pm$0.10     & 28.6$\pm$0.2    & 0.1243$\pm$0.0042 & 83.3$\pm$2.8       & 8.917$\pm$0.006      & 8.1  $\pm$0.3      & 0.657  \\
AB04    & 46.83   & 3.25$\pm$0.10     & 30.5$\pm$0.4    & 0.1063$\pm$0.0028 & 71.2$\pm$1.9       & 8.349$\pm$0.020      & 9.4  $\pm$0.3       & 0.920 \\
AB05    & 41.27   & 3.56$\pm$0.06     & 28.6$\pm$0.4    & 0.1244$\pm$0.0034 & 83.4$\pm$2.2       & 9.401$\pm$0.009      & 8.1  $\pm$0.2       & 1.315 \\
AB06    & 38.21   & 4.35$\pm$0.23     & 26.7$\pm$0.1    & 0.1633$\pm$0.0087 & 109.4$\pm$5.8      & 10.760$\pm$0.005     & 6.2  $\pm$0.3       & 1.534 \\
AB07    & 32.30   & 2.92$\pm$0.15     & 29.1$\pm$0.4    & 0.1003$\pm$0.0049 & 67.2$\pm$3.3       & 9.159$\pm$0.017      & 10.1 $\pm$0.5       & 1.972 \\
AB08    & 27.80   & 3.84$\pm$0.13     & 27.9$\pm$0.4    & 0.1379$\pm$0.0049 & 92.4$\pm$3.3       & 9.494$\pm$0.003      & 7.3  $\pm$0.3        & 1.972\\
AB09    & 23.48   & 3.61$\pm$0.09     & 28.6$\pm$0.1    & 0.1264$\pm$0.0027 & 84.7$\pm$1.8       & 8.126$\pm$0.007      & 7.9  $\pm$0.2        & 1.657\\
AB10    & 17.90   & 3.10$\pm$0.09     & 29.4$\pm$0.3    & 0.1054$\pm$0.0027 & 70.6$\pm$1.8       & 7.494$\pm$0.004      & 9.5  $\pm$0.2        & 1.255\\
AB11    & 13.17   & 2.55$\pm$0.15     & 30.7$\pm$0.2    & 0.0830$\pm$0.0048 & 55.6$\pm$3.2       & 6.105$\pm$0.018      & 12.2 $\pm$0.7        & 0.920\\
AB12    & 6.40   & 1.13$\pm$0.10     & 32.7$\pm$0.3    & 0.0348$\pm$0.0030 & 23.3$\pm$2.0       & 3.478$\pm$0.009      & 30.0 $\pm$3.0        & 0.445\\
AB13    & 0.96   & 0.21$\pm$0.07     & 33.5$\pm$0.2    & 0.0062$\pm$0.0021 & 4.2 $\pm$1.4       & 0.574$\pm$0.001      & 319.0$\pm$112.2       & 0.067  \\
\hline
\hline
\end{tabular}
\end{table*}

\begin{figure}[htb]
\centering
\includegraphics[width=1\linewidth]{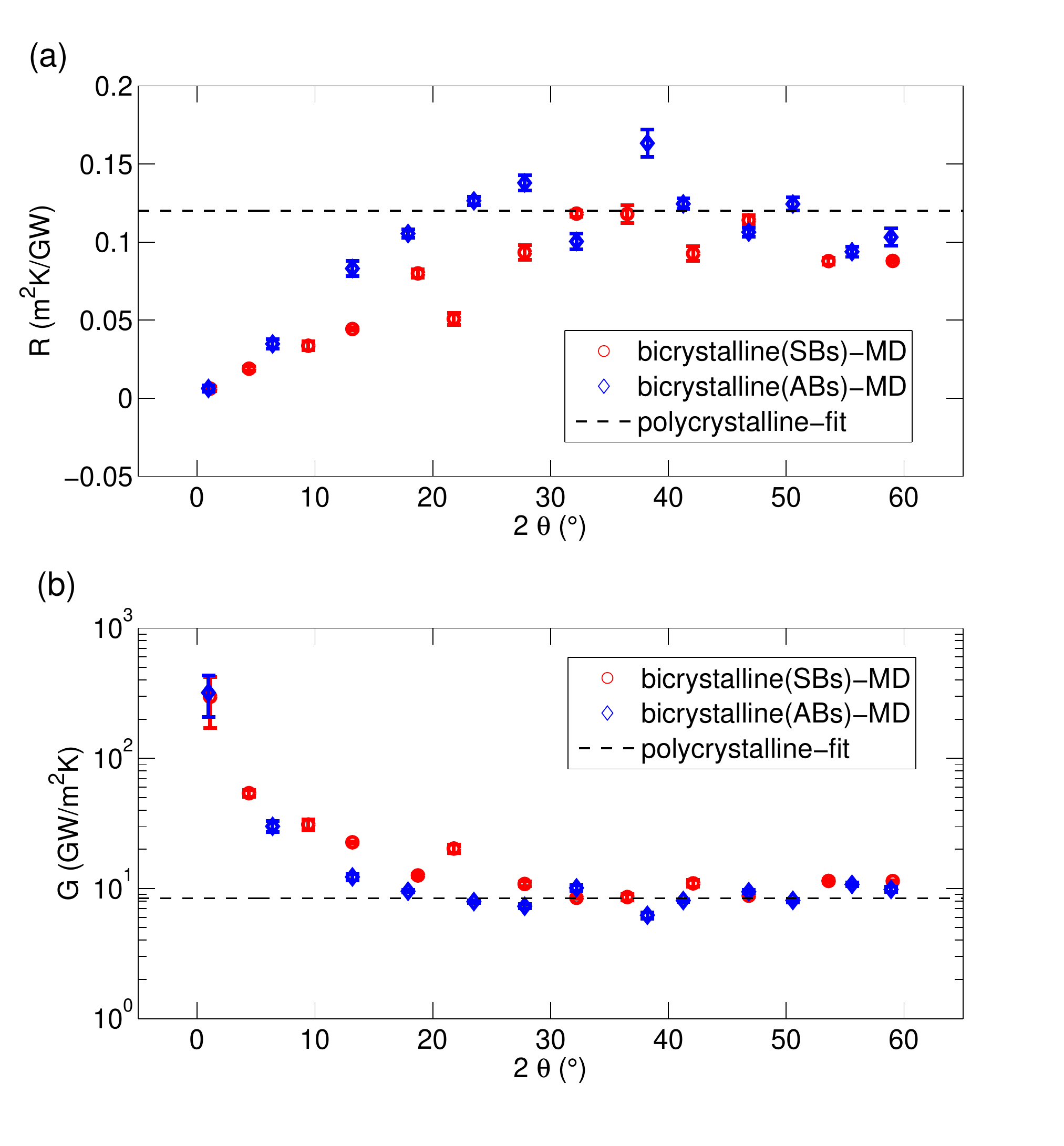}
\caption{(a) Kapitza resistance $R$ and (b) Kapitza conductance $G$ as a function of the tilt angle $2\theta$ of the GBs calculated from the NEMD simulations.The dashed lines represent the fitted values for the polycrystalline h-BN samples.}
\label{fig_kr}
\end{figure}

\begin{figure}[htb]
\centering
\includegraphics[width=1\linewidth]{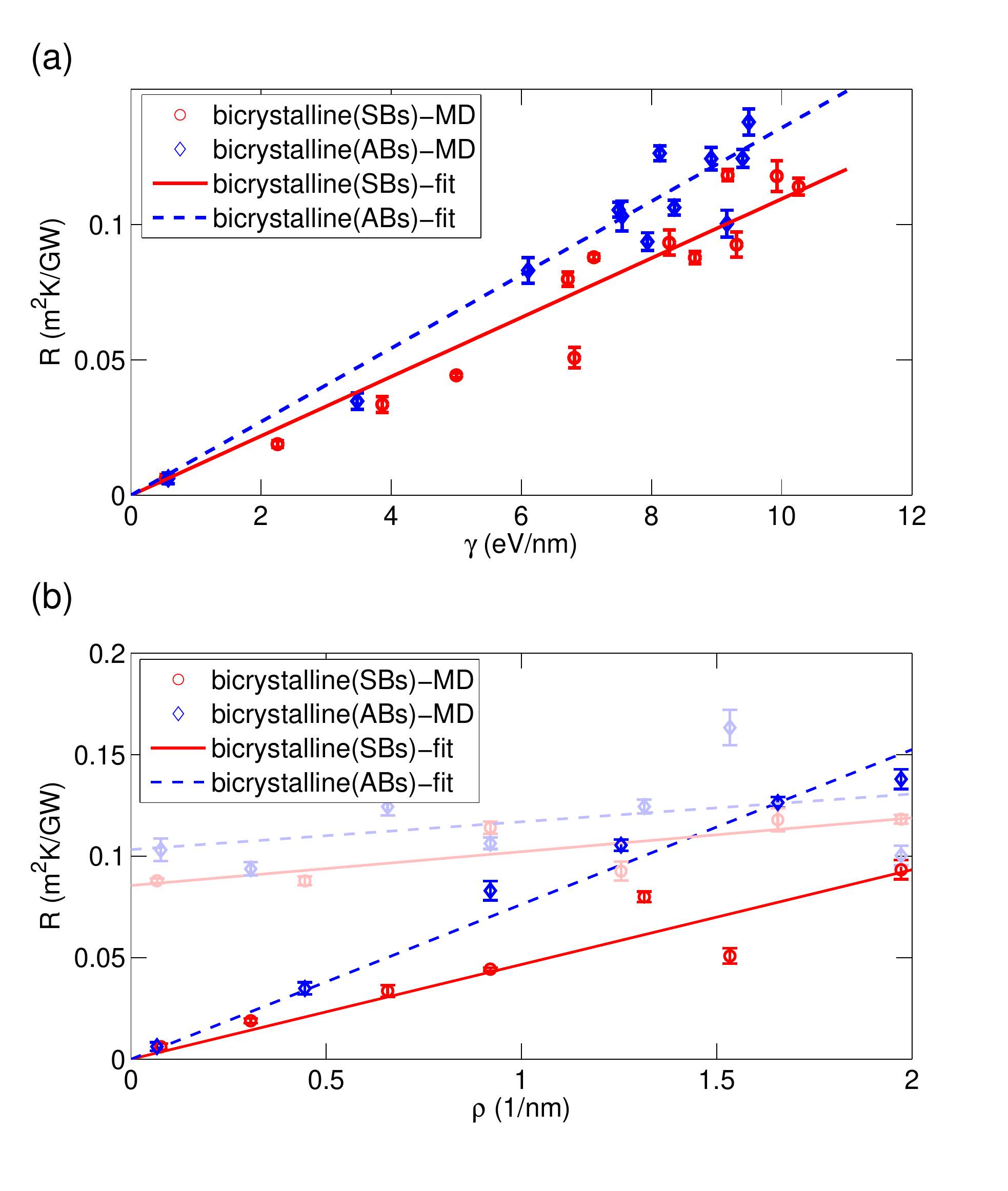}
\caption{(a) Kapitza resistance $R$ as a function of the GB line tension $\gamma$. The markers are results from NEMD simulations and the lines are linear fits according to $R = c\gamma$, with $c=0.011$ and $0.014$ (when $R$ is in units of m$^2$K/GW and $\gamma$ is in units of eV/nm) for SBs and ABs, respectively. (b)  Kapitza resistance $R$ versus the dislocation density $\rho$. In (b), the full color markers and lines correspond to $2\theta < 30^\circ$ and the fainter ghost markers and lines to $2\theta > 30^\circ$. }
\label{fig_RE}
\end{figure}

We next consider heat transport properties of individual grain boundaries. The temperature jump $\Delta T$, heat flux $Q$, Kapitza resistance $R$, and Kapitza conductance $G$ for the $26$ GBs (13 SBs and 13 ABs) calculated using the NEMD method are presented in Table \ref{tab_02}. To understand the results, we plot in Fig. \ref{fig_kr} $R$ and $G$ against the the tilt angle $2\theta$ of the GBs. It is obvious that the Kapitza resistance $R$ depends strongly on the tilt angle: it first increases almost linearly with increasing tilt angle and then saturates in the intermediate-angle region. 

The relationship between the Kapitza resistance and the atomistic structure of the GBs can be better appreciated by considering the GB line tension (grain boundary energy density) $\gamma$, which is defined as
\begin{equation}
\gamma = \lim_{L_y \to \infty} \frac{\Delta E}{L_y},
\end{equation}
where $\Delta E$ is the formation energy of a GB with length $L_y$. Figure \ref{fig_RE} shows that the Kapitza resistance is almost linearly proportional to the line tension and the fitted proportionality constant in the case of ABs is larger than that in the case of SBs. The difference in the proportionality constant may be related to the larger mass disorder in the ABs which does not contribute to the system energy as the Tersoff potential for h-BN \cite{sevik2011prb} we adopted does not distinguish the boron and nitrogen atoms.

Figure \ref{fig_RE} (b) plots the Kapitza resistances $R$ of the SBs and ABs as a function of the dislocation density $\rho$. Since some GB structures contain multiple dislocation types with different sizes and different Burgers vectors, we define $\rho$ as follows. We assume all dislocations have a Burgers vector equal to the lattice constant and perpendicular to the GBs. The dislocation density $\rho$ is then the number of such dislocations required per unit length along a GB of a particular tilt angle. This definition is trivial when $2\theta < 30^\circ$ and where simple chains of dislocations are encountered, but for $2\theta > 30^\circ$ it is not. Here, inversion boundaries at $2\theta = 60^\circ$ are approached and, while $\rho$ again declines, eventually all the way back to zero, defects with a zero-Burgers vector take the dislocations' place. It then becomes very difficult to define what constitutes a single defect.

Following the above definition for the dislocation density $\rho$, we find almost linear scaling for the Kapitza resistance $R$ as its function when $2\theta < 30^\circ$. This makes sense as the density of phonon-scattering dislocations increases monotonically. Abnormal data points are observed for structures SB05 and SB06 at $\rho \approx 1.3$ 1/nm and $\rho \approx 1.5$ 1/nm corresponding to $2\theta \approx 18.7^\circ$ and $2\theta \approx 21.8^\circ$, respectively. The former shows a higher $R$ most likely due to GBs composed of a mixture of three different dislocation types. The latter case corresponds to a well-known \cite{yazyev2010prb, hirvonen2016prb, azizi2017cb} high-symmetry boundary between tilted honeycomb lattices. Low $R$ is due to a flat GB \cite{azizi2017cb}. For $2\theta > 30^\circ$, it is not clear what kind of behavior is to be expected. It appears that $R$ decreases very slightly as more and more dislocations are replaced by zero-Burgers vector defects when inversion boundaries are approached. Here, an almost-constant $R$ makes sense as the corresponding GBs display continuous chains of defects --- only their type varies with the tilt angle.

It is interesting to compare the results for h-BN GBs with those for graphene GBs. The largest Kapitza resistance of h-BN GBs occurs in the AB06 system, being about $0.16$ m$^2$K/GW, which is about three times as large as the largest Kapitza resistance of graphene GBs \cite{azizi2017cb} (about $0.06$ m$^2$K/GW). A more reasonable comparison between different materials is in terms of the Kapitza length \cite{nan1997jap} $L_{\rm K}$, which is defined such that the thermal resistance of the pristine material of length $L_{\rm K}$ equals the Kapitza resistance, i.e.,
\begin{equation}
L_{\rm K}/\kappa_0 = R.
\end{equation}
The Kapitza lengths calculated from the Kapitza resistances and the pristine h-BN thermal conductivity $\kappa_0=670$ W/mK calculated above are listed in Table \ref{tab_02}. The largest Kapitza lengths in h-BN and graphene GBs are about $110$ nm and $180$ nm, respectively. We see that although the h-BN GBs have larger Kapitza resistances, they have shorter Kapitza lengths $L_{\rm K}$ as compared to the graphene GBs.

\subsection{Thermal conductivity of polycrystalline h-BN}

\begin{figure}[htb]
\centering
\includegraphics[width=1\linewidth]{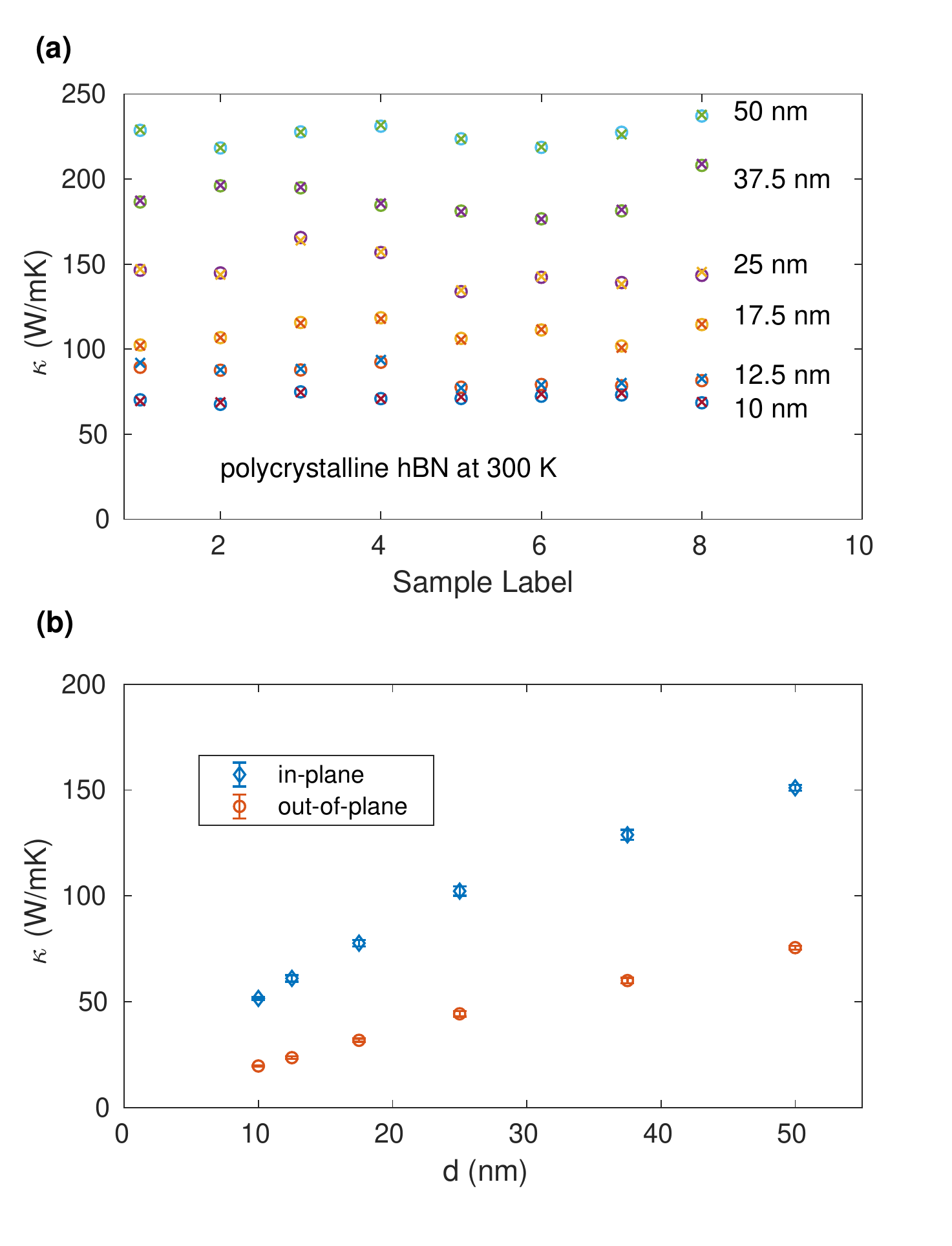}
\caption{(a) Thermal conductivity of polycrystalline h-BN at 300 K calculated using the HNEMD method for different grain sizes (from $d=10$ nm to $d=50$ nm) and PFC realizations (``Sample Label''). The circles and the crosses represent results from two independent runs. (b) In-plane and out-of-plane thermal conductivity components of polycrystalline h-BN at 300 K as a function of grain size $d$. The error bars are calculated as the standard error from the eight samples for each grain size.  }
\label{fig_poly_kappa_all}
\end{figure}

\begin{figure}[ht]
\centering
\includegraphics[width=1\linewidth]{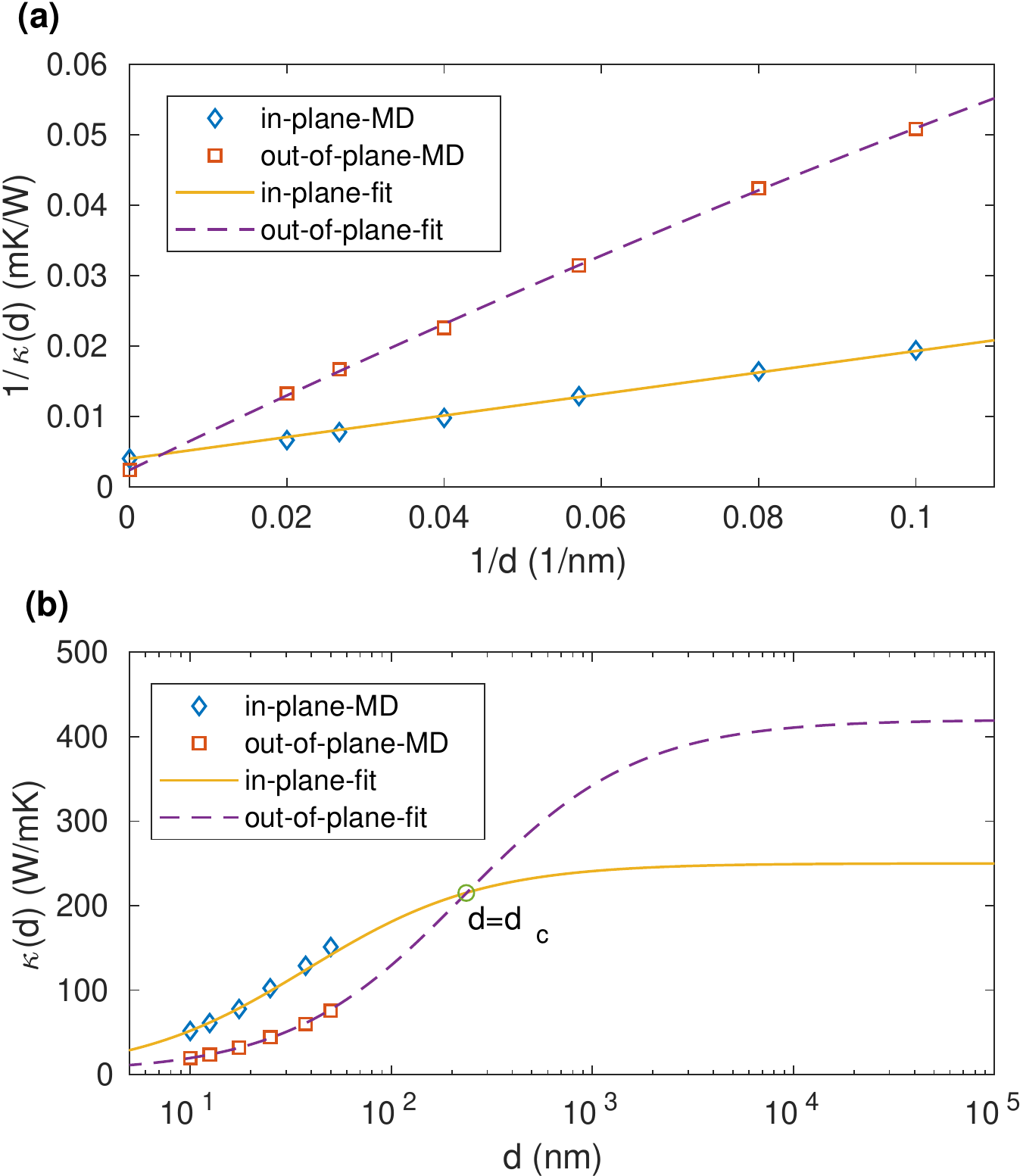}
\caption{(a) Inverse thermal conductivity $1/\kappa(d)$ as a function of the inverse grain size $1/d$ for polycrystalline h-BN at 300 K. The in-plane and out-of-plane components are fitted using a linear and a quadratic function, respectively. (b) The same data as in (a) but for $\kappa(d)$ as a function of $d$ in logarithmic scale.}
\label{fig_poly_kappa_fit}
\end{figure}

Last, we consider heat transport in polycrystalline h-BN. The thermal conductivity values for all the polycrystalline samples calculated using the HNEMD method are shown in Fig. \ref{fig_poly_kappa_all}(a). For each grain size $d$, there are eight samples, and for each sample we have performed two independent MD simulations. The results from the two independent simulations are very close to each other, indicating the high accuracy of the HNEMD method. We have also tried to use the EMD method, which turned out to be much less efficient. After averaging over the two independent simulations, we get eight thermal conductivity values for each grain size. From these, we obtain an average and an error estimate for each grain size and the results are shown in Fig. \ref{fig_poly_kappa_all}(b). Again, we have decomposed the total thermal conductivity into an in-plane component and an out-of-plane component. For both components, the thermal conductivity increases with increasing grain size, but the out-of-plane component has lower thermal conductivity, which is opposite to the case of pristine h-BN. This means that the two components have quite different scalings of the thermal conductivity with respect to the grain size, as we discuss below. 
 
The scaling of the thermal conductivity $\kappa(d)$ in polycrystalline h-BN with respect to the grain size $d$ can be better understood by considering the inverse thermal conductivity $1/\kappa(d)$ as a function of the inverse grain size $1/d$, as shown in Fig. \ref{fig_poly_kappa_fit}(a). The grain size of the pristine h-BN is chosen as infinity by definition. If we regard the total thermal resistance $d/\kappa(d)$ in polycrystalline h-BN as a sum of that from the grains $d/\kappa_0$ and the GBs $1/G$, we can write down the following series resistance formula:
\begin{equation}
\label{equation:kappa_d}
\frac{1}{\kappa(d)} = \frac{1}{\kappa_0} + \frac{1}{Gd}.
\end{equation}
This is just an approximate relation because it does not consider the frequency dependence of the thermal conductivity. Actually, this formula is closely related to the ballistic-to-diffusive transition formula \cite{schelling2002prb} used in the finite-size scaling analyses of NEMD simulations:
\begin{equation}
\frac{1}{\kappa(L)} = \frac{1}{\kappa_0} \left(1 + \frac{\lambda}{L}\right),
\end{equation}
where $\lambda$ is the effective phonon mean free path and $\kappa(L)$ is the thermal conductivity of a system with length $L$, which is the source-sink distance in the NEMD simulation. To see the connection between the two equations above, we note that $\kappa_0$ is related to the ballistic conductance $G_0$ by $\kappa_0=G_0\lambda$. Using this, we can rewrite the ballistic-to-diffusive formula as
\begin{equation}
\label{equation:kappa_L}
\frac{1}{\kappa(L)} = \frac{1}{\kappa_0} + \frac{1}{G_0L}.
\end{equation}
Comparing Eqs. (\ref{equation:kappa_d}) and (\ref{equation:kappa_L}), we can identify the analogies between $d$ and $L$ and between $G$ and $G_0$. The applicability of Eq. (\ref{equation:kappa_L}) has been studied in previous works \cite{dong2018prb,sellan2010prb} and it was found that the linear relationship between $1/\kappa(L)$ and $1/L$ is only valid when $L$ is comparable or larger than $\lambda$ such that $\kappa(L)$ is close to $\kappa_0$. Due to the above analogy, the applicability of Eq. (\ref{equation:kappa_d}) is similar.  For the in-plane component, where $\kappa^{\rm in}(d)$ are close to $\kappa_0^{\rm in}$, we find that Eq. (\ref{equation:kappa_d}) holds well:
\begin{equation}
\label{equation:kappa_d_in}
\frac{1}{\kappa^{\rm in}(d)} = \frac{1}{\kappa_0^{\rm in}} + \frac{1}{G^{\rm in}d},
\end{equation}
where the Kapitza conductance from the in-plane phonons is fitted to $G^{\rm in}=6.5$ GW/m$^{2}$K. For the out-of-plane component, where $\kappa^{\rm out}(d)$ are much smaller than $\kappa_0^{\rm out}$, we find that a quadratic scaling describes the data better:
\begin{equation}
\frac{1}{\kappa^{\rm out}(d)} = \frac{1}{\kappa_0^{\rm out}}
+ \frac{1}{G^{\rm out}d} \left(1 + \frac{\alpha}{d}\right),
\end{equation}
with $G^{\rm out}=1.9$ GW/m$^{2}$K and $\alpha=-1.0$ nm. The total Kapitza conductance is thus $G=G^{\rm in} + G^{\rm out} = 8.4$ GW/m$^{2}$K and the total Kapitza resistance is $R=1/G=0.12$ m$^{2}$K/GW. The fitted $R$ and $G$ values here are plotted as dashed lines in Fig. \ref{fig_kr}. We see that the effective Kapitza resistance/conductance in polycrystalline h-BN is comparable to those for the individual GBs in bicrystalline h-BN with large ($2\theta>30 ^\circ$) tilt angles. 

Because $G^{\rm in} > G^{\rm out}$ and $\kappa_0^{\rm in} < \kappa_0^{\rm out}$, there must be a crossover point for the grain size $d_{\rm c}$ [Fig. \ref{fig_poly_kappa_fit}(b)]: when $d<d_{\rm c}$, $\kappa^{\rm in}(d) > \kappa^{\rm out}(d)$ and when $d>d_{\rm c}$, $\kappa^{\rm in}(d) < \kappa^{\rm out}(d)$. This bimodal grain size scaling of the thermal conductivity is similar to the case of suspended polycrystalline graphene. The origin of this bimodal grain size scaling is the large difference between the average phonon mean free paths of the in-plane and the out-of-plane phonons in pristine systems. For pristine h-BN, $\lambda^{\rm in}=\kappa^{\rm in}/G^{\rm in}\approx 38$ nm and $\lambda^{\rm out}=\kappa^{\rm out}/G^{\rm out}\approx 220$ nm.

\section{Summary and Conclusions}

In the present work we have employed extensive nonequilibrium and homogeneous nonequilibrium MD simulations with a Tersoff potential \cite{sevik2011prb} to study heat transport in pristine, bicrystalline, and polycrystalline single-layer h-BN systems. For isotopically pure and pristine h-BN, we have obtained a thermal conductivity of $\kappa_0=670\pm 30$ W/mK, which is dominated by the flexural phonons. Isotope scattering reduces $\kappa_0$ to $520 \pm 30$ W/mK, which is comparable to the experimental value for double-layer h-BN \cite{wang2016sr}.

For systems with grain boundaries we have used the two-component PFC model to generate large relaxed samples. We have calculated the Kapitza thermal resistance $R$ across 26 individual grain boundaries with the tilt angles $2\theta$ ranging from 0 to $60^\circ$. The antisymmetric grain boundaries have larger $R$ on average as compared to the symmetric ones. For each grain boundary type, $R$ is found to be almost linearly proportional to the grain boundary line tension in the whole range of tilt angle and to be linearly proportional to the grain boundary defect density only in the range of $2\theta < 30^\circ$. For multigrain h-BN systems, the scaling of the thermal conductivity exhibits a bimodal behavior: the thermal conductivity from the in-plane phonons converges much faster than that of the out-of-plane phonons with increasing grain size. 
The total Kapitza conductance ($8.4$ GW/m$^{2}$K) is comparable to those of individual grain boundaries with large ($2\theta>30^\circ$) tilt angles. 

Finally, we would like to note that we have not considered quantum effects in our classical MD simulations here. Quantum corrections based on spectral decomposition could be made for $R$ following Refs. \cite{azizi2017cb,fan2017nl}, where $R$ for graphene at 300 K was reduced by a factor of $2.5$ due to the high Debye temperature $T_{\rm D} \approx 2000$ K. Unfortunately, there is so far no \cite{turney2009prb,bedoya2014prb} reliable quantum correction method for the thermal conductivity $\kappa_0$ of pristine systems in the diffusive transport regime. Also, we are not aware of any experimental results on $T_{\rm D}$ for two-dimensional h-BN, but we have computed it for the empirical Tersoff potential used in our simulations and find that $T_{\rm D} = 1740$ K at 300 K. Based on this we can estimate that quantum statistical corrections should reduce $R$ approximately by a factor of two from the values reported based on our MD simulations.

\textbf{Note}: All the bicrystalline samples can be found as a supplementary from journal (PCCP) and all the polycrystalline samples can be found from Zenodo: \url{http://doi.org/10.5281/zenodo.1400276}.

\begin{acknowledgments}
This work was supported by the National Natural Science Foundation of China under Grant No. 11404033, the Natural Science Foundation of Liaoning Province under Grant No. 20180550102, and the Academy of Finland QTF Centre of Excellence program (Project 312298). P.H. acknowledges financial support from the Vilho, Yrj\"o and Kalle V\"ais\"al\"a Foundation of the Finnish Academy  of  Science  and  Letters. We acknowledge the computational resources provided by Aalto Science-IT project and Finland's IT Center for Science (CSC).
\end{acknowledgments}

\end{document}